\documentclass{evn2004}
\usepackage{txfonts}
\usepackage{psfig}
\usepackage{graphicx}

\begin{document}
 
\setcounter{page}{213}

   \title{mm VLBA Observations of SiO masers: probing the close stellar
          environment of the PPN OH231.8+4.2}

   \author{Desmurs, J.-F.\inst{1} \and S\'anchez Contreras, C.\inst{2}
          \and Bujarrabal, V.\inst{3} \and Alcolea, J.\inst{1} \and
          Colomer, F.\inst{3}}

   \institute{Observatorio Astron\'omico Nacional, c/Alfonso XII 3,
     28014 Madrid, Spain \and California Institute of Technology,
     Astronomy Department, Pasadena, California, USA. \and OAN,
     Apartado 112, E-28803 Madrid, Spain}

   \abstract{ We present milliarsecond-resolution maps of the SiO maser
emission $v$=1 J=2--1 (3~mm) in the bipolar post-AGB nebula
OH\,231.8+4.2, and compare them with our previous observations of the
$v$=2 J=1--0 line (7~mm). Our observations show that the SiO masers
arise in several bright spots forming a structure elongated in a
direction perpendicular to the symmetry axis of the nebula. This, and
the complex velocity gradient observed, is consistent with the presence
of an equatorial torus in rotation and with an infall of material
towards the star.}

   \authorrunning{J.-F. Desmurs et al.}
   \titlerunning{mm-VLBI observations of SiO in OH231.8+4.2}

\maketitle


\section{Introduction}

Planetary and Proto-Planetary Nebulae (PNe, PPNe) present conspicuous
departures from spherical symmetry, including e.g. multiple lobes and
jets. To explain their evolution from spherical AGB envelopes, several
models have postulated the presence of dense rings or disks close to
the central post-AGB stars as the agents of the mechanical collimation
of the stellar wind. Existing observations reveal the presence of
central disks in several PPNe, but their limited spatial resolution
cannot unveil the very inner regions of the disks that are relevant
for the processes mentioned above.

Our first VLBA observations of SiO masers in OH\,231.8+4.2, carried
out at 7~mm ($v$=2, $J$=1--0) in April 2000, revealed for the first
time the structure and kinematics of the close stellar environment in
a PPN (Sanchez Contreras et al., 2002). The SiO maser emission arises
in several compact, bright spots forming a structure elongated in the
direction perpendicular to the symmetry axis of the nebula. Such a
distribution is consistent with an equatorial torus with a radius of
$\sim$\,6~AU around the central star. A complex velocity gradient was
found along the torus, which suggests rotation and infall of material
towards the star. The rotation and infalling velocities deduced are of
the same order and range between $\sim$\,7 and
$\sim$\,10\,km\,s$^{-1}$.  Such a distribution is remarkably different
from that found in other late-type stars, where the masers form a
roughly spherical ring-like chain of spots resulting from tangential
maser amplification in a thin, spherical shell (Diamond et al., 1994,
Desmurs et al., 2000).

\section{Observations and Results}

\begin{figure}
  \centering \includegraphics[angle=-90, width=0.7\textwidth]{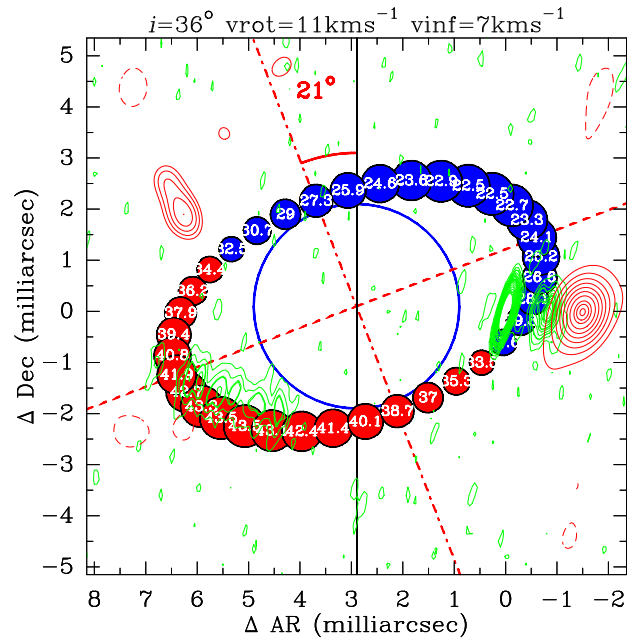}
  \caption {Comparison of our first observations of the $v$=2, $J$=1--0
(7~mm) SiO maser transitions (grey scale+contours), with our new
observations of the $v$=1, $J$=2--1 (3~mm) SiO maser (thick contours)
and our model of a rotating/infalling torus in the OH\,231.8+4.2 (grey
\& light grey circles; the velocity of the spots predicted by the model
is indicated inside the circles --- see Sanchez Contreras et al., 2002
for details).}
\end{figure} 

The new VLBA observations at 3~mm were observed on April 2002. The
data were recorded in dual circular polarization over a bandwidth of
16~MHz, assuming a rest frequency of 86243.442 MHz for the $v$=1
$J$=2--1 SiO maser emission, for a final spectral resolution of about
0.1 km s$^{-1}$.  We frequently observed several calibrators for
band-pass and phase calibration; pointing was checked and corrected
every less than half an hour. The data reduction followed the standard
scheme for spectral line calibration data in AIPS. To improve the
phase calibration solutions, we extensively map all the calibrators,
and their brightness distribution were introduced and taken into
account in the calibration process.

Our VLBA maps at 3~mm (Fig. 1) show two SiO maser spots that are
globally aligned with those at 7~mm, i.e., they are roughly
distributed orthogonally to the nebular symmetry axis.  The velocities
of the two maser spots observed at 3~mm are in good agreement with
velocities observed at 7~mm. The spatial distribution and velocity
structure of the SiO maser spots at 3~mm and 7~mm are consistent with a
torus with radius $R \sim$\,6\,A.U$.$, perpendicular to the nebular
symmetry axis (inclined $\sim36^o$ with respect to the plane of the
sky). This torus-like structure is not totally traced by the maser
spots.  Such a spot distribution is expected, even for a homogeneous
torus, if the maser is tangentially amplified. This amplification
mechanism leads to the most intense maser features at the edges of the
torus-like structure. Moreover, given the comparable size of the star
and the maser emitting region, occultation by the star of the far side
of the torus is very likely, which would partially explain the small
number of detected spots.

In our opinion, the structure traced by the SiO maser emission in
OH\,231.8+4.2, compatible with a rotating+infalling torus very close
to the central star, is unlikely an accretion disk. The dynamic is
well reproduced by a simple model considering the presence of an inner
torus in rotation and infall with a radius of $\sim$~5AU (R$_{star}
\sim$4.5~AU), which means very close to the surface of the star
(Sanchez Contreras et al., 2002).  In fact, the dramatic changes with
time of the SiO $v$=2, $J$=1--0 profile suggest that the masers lie in
a region with an unstable structure and/or kinematics, rather than in
a stable accretion disk. Moreover, accretion disks are expected around
the compact companion in a binary system and less likely around the
mass-losing star.

Finally, it is worth mentioning that: (a) 3~mm maser spots are located
further away from the central star than those at 7~mm, similarly to
what has been recently found in AGB stars (see Soria-Ruiz et al.,
2004); and (b) the rotation velocity measured in the torus of
OH\,231.8+4.2 is consistent with the values of the rotation velocity
found in inner envelopes of several AGBs (NML Cyg : Boboltz \& Marvel,
2000; TX CAM and IRC+10011: Desmurs et al., 2001, R Aqr : Hollis et
al., 2001).

\end{document}